\documentclass[prl,twocolumn,showpacs,showkeys]{revtex4} 

\usepackage{epsfig}
\usepackage{graphicx}

\DeclareMathAlphabet{\mathitb}{OT1}{cmr}{bx}{sl}
\begin{document}
\title{Direct observation of the Aharonov-Casher phase}

\date{\today}
\author{M.~K\"{o}nig$^1$}
\author{A.~Tschetschetkin$^1$}
\author{E.~M.~Hankiewicz$^{2,3}$}
\author{Jairo~Sinova$^2$}
\author{V.~Hock$^1$}
\author{V.~Daumer$^1$}
\author{M.~Sch\"{a}fer$^1$}
\author{C.~R.~Becker$^1$}
\author{H.~Buhmann$^1$}
\author{L.~W.~Molenkamp$^1$}
\affiliation{$^1$Physikalisches Institut(EP 3), Universit\"{a}t
W\"{u}rzburg, Am Hubland, 97074 W\"{u}rzburg, Germany\\
$^2$Department of Physics, Texas A\&M University, College Station,
Texas 77843-4242, USA\\ $^3$ Department of Physics and Astronomy
University of Missouri, Columbia, Missouri 65211, USA}

\begin{abstract}
Ring structures fabricated from HgTe/HgCdTe quantum wells have
been used to study Aharonov-Bohm type conductance oscillations as
a function of Rashba spin-orbit splitting strength. We observe
non-monotonic phase changes indicating that an additional phase
factor modifies the electron wave function. We associate these
observations with the Aharonov-Casher effect. This is confirmed
by comparison with numerical calculations of the
magneto-conductance for a multichannel ring structure within the
Landauer-B\"uttiker formalism.
\end{abstract}



\maketitle

In the early 1980s it was shown that a quantum mechanical system
acquires a geometric phase for a cyclic motion in parameter
space. This geometric phase under adiabatic motion is called
Berry phase \cite{Berry84}, while its later generalization to
include non-adiabatic motion is known as Aharonov-Anandan phase
\cite{Aharonov87}. A manifestation of the Berry phase is the well
known Aharonov-Bohm (AB) phase \cite{Aharonov59} of an electrical
charge which cycles around a magnetic flux. Aside from the AB
effect, the first experimental observation of the Berry phase was
reported in 1986 for photons in a wound optical fiber
\cite{Tomita86}. Another important Berry phase effect is the
Aharonov-Casher (AC) effect \cite{Aharonov84}, which has been
proposed to occur when an electron propagates in a ring structure
in an external magnetic field perpendicular to the ring plane in
the presence of {\sc SO} interaction \cite{Nitta99}.

This AC effect can be seen when two partial waves move around the
ring in different directions. They will acquire a phase
difference which depends on the spin orientation with respect to
the total magnetic field
$\vec{B}_{tot}=\vec{B}_{ext}+\vec{B}_{eff}$ and the path of each
partial wave. $\vec{B}_{eff}$ is the effective field induced by
the SO interaction. The phase difference is approximately
\cite{Nitta99}
\begin{eqnarray}
\Delta\varphi_{\psi_{s}^+ - \psi_{s}^-} & =
-2\pi\frac{\Phi}{\Phi_0}-b\pi(1-cos\theta)\label{upup}\\
\Delta\varphi_{\psi_{s}^+ - \psi_{\bar{s}}^-} & =
-2\pi\frac{\Phi}{\Phi_0}-b 2\pi r\frac{m^{\ast}\alpha}{\hbar^2}sin\theta\label{updown}
\end{eqnarray}
where $s=\uparrow$ and $\downarrow$ denote parallel and
anti-parallel orientation to $\vec{B}_{tot}$, $b=+1$ for
$s=\uparrow$ and $b=-1$ for $s=\downarrow$, and the superscript
$-$($+$) denotes a clockwise (counterclockwise) evolution,
respectively. In the above equations, $\alpha$ is the SO
parameter, $r$ the ring radius, $m^{\ast}$ the effective electron
mass and $\theta$ the angle between the external
($\vec{B}_{ext}$) and the total magnetic field $\vec{B}_{tot}$.
For both equations, the first term on the right hand side can be
identified with the AB phase and the second term of
Eq.~(\ref{upup}) with the geometric Berry or Aharonov-Anandan
phase. The second term in Eq.~(\ref{updown})
 represents the dynamic part of the AC
phase, i.e. the phase of a particle with a magnetic moment that
moves around an electric field. From the expressions above, it can
be seen that an increase of the AC phase will lead to a phase change
that increases continuously with $\alpha$, whereas the contribution
due to the geometric phase results in a phase shift limited to
$\Delta\varphi_{geom} \leq \pi$.

Both the AC phase \cite{Mathur92} and the geometric phase
\cite{Aronov93,Qian94} depend on the {\sc SO} interaction. As a
result, one expects a complicated non-monotonic interference
pattern as a function of magnetic field and {\sc SO} interaction
strength. So far, to our knowledge, apart  from the AB effect no
direct observation of phase related effects in solid state
systems has been reported. Recently, side bands in the Fourier
transform of AB oscillations have been interpreted as an
indication for the existence of a Berry phase
\cite{Morpurgo98,Yau02,Yang04}. However, these interpretations
have been questioned \cite{Deraedt99,Malshukov03,Wagh03}. Spin
interference signals in square loop arrays have been recently
reported by Koga {\it et al.}~\cite{Koga05} but their direct
relationship to the AC effect is not easily established. It thus
would be important to observe these phase-related effects
directly.

Here, we present experimental results on the magneto-transport
properties of a HgTe ring structure. The strength of the Rashba
effect was controlled via a gate electrode by varying the
asymmetry of the quantum well structure. We observe systematic
variations in the conductance of the device as a function of both
external B-field and gate voltage. The gate-voltage dependent
oscillations clearly exhibit a non-monotonic phase change, which
is related to the dynamic part of the AC phase. This
interpretation is confirmed by numerical calculations for
multichannel rings within the Landauer-B\"uttiker formalism.

\begin{figure}[t]
\epsfig{figure=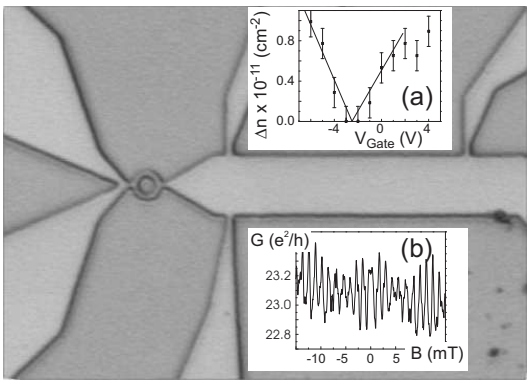,width=2.75in}

\caption{Microscope image of the sample layout without top gate
electrode. The average ring radius is 1 $\mu$m and the width of
each arm is 300 nm. Inset (a): The symmetry point is estimated
from the subband population differences as a function of the
applied gate voltage, i.e., $V_{Gate}\approx -2.57$~V. Inset (b)
shows the conductance oscillations at $V_{Gate}=-2.57$~V. The
period, $\Delta B = 1.1~{\rm mT}$, is consistent with the ring
radius.} \label{sample}
\end{figure}

The samples are based on type-III HgTe/HgCdTe quantum well (QW)
structures with electron mobilities of the order of
$10^5$~cm${^2}$/(Vs). This narrow-gap material exhibits a strong
Rashba spin-orbit ({\sc SO}) splitting \cite{Rashba60}, which can
be modified over a wide range via an externally applied gate
voltage \cite{Zhang01,Gui04}. The {\it n}-type QWs are
symmetrically modulation doped and have been epitaxially grown in
a MBE system \cite{Zhang01,Gosch98}. The width of the HgTe-QW is
12~nm. The device structure is fabricated by optical and e-beam
lithography and wet chemical etching \cite{Daumer03}. A picture
of the sample is shown in Fig.~\ref{sample}. The width of leads
in each arm of the ring is 300~nm and the ring radius is
1~$\mu$m. Directly attached to the ring is a Hall bar. Both
components, ring and Hall bar, are covered by an insulator
(Si$_3$N$_4$) and a metal gate electrode (Au). Because of these
layers, the structure becomes asymmetric and a non-zero Rashba
splitting is found for $V_{Gate}=0$. For all measurements, the
samples were mounted in a $^3$He/$^4$He dilution cryostat with a
base temperature of $\sim 20$~mK.

\begin{figure}[t]
\epsfig{figure=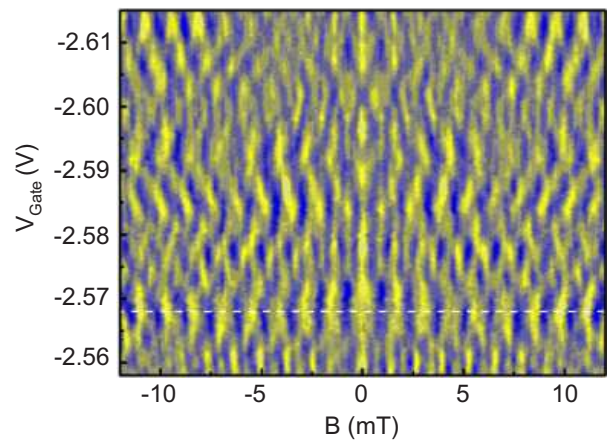,width=3.1in}

\caption{Color plot of the conductance fluctuations, maxima
(yellow) and minima (blue), as a function of magnetic field and
gate voltage. The dashed line indicates $\Delta_{R}=0$ which can
be located at $V_{Gate}=-2.568$~V, based on the mirror symmetry
of the conduction around this line.} \label{2dtotal}
\end{figure}

\begin{figure}[h]
\epsfig{figure=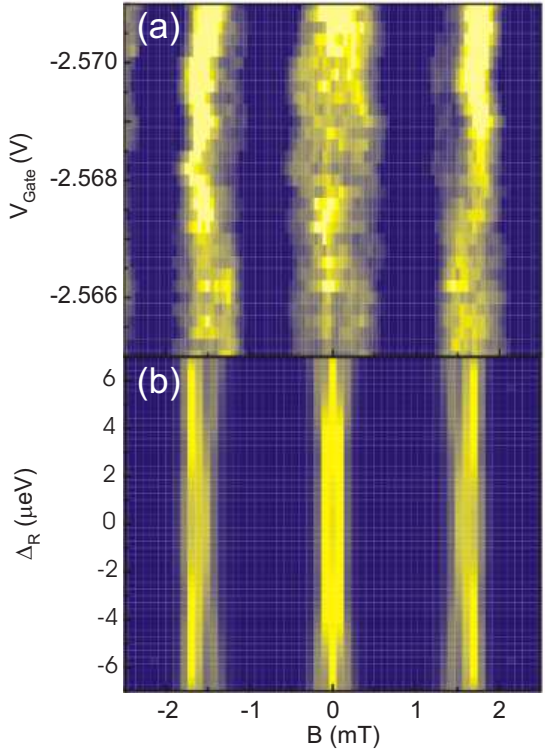,width=2.9in}

\caption{(a) Near the symmetry point ($V_{Gate}=-2.568$~V) the
interference pattern is unperturbed by the {\sc SO} interaction.
(b) The theoretical calculations for a 6-channel ring show
consistent results for the corresponding range of $\Delta_{R}$.
Yellow and blue correspond to conductance maxima and minima,
respectively.} \label{B0}
\end{figure}

An applied gate voltage leads to a change in carrier
concentration, electron mobility and the Rashba {\sc SO}
splitting energy ($\Delta_{R}$). These sample parameters were
deduced directly from magneto-transport measurements of the
attached Hall bar structure (see Fig.~\ref{sample}). The Rashba
SO splitting energy of the presented sample could be varied from
zero up to $\Delta_{R} \approx 6$~meV. These values are obtained
by analyzing the Fourier transform of the Shubnikov-de Haas (SdH)
oscillations. The Rashba splitting energy depends linearly on the
applied gate voltage for $0$~V $\leq V_{Gate} \le-4$~V.
 This allows an estimation of the symmetry point for
$\Delta_{R} = 0$ at $V_{Gate} \approx (-2.57\pm 0.02)$~V
[Fig.~\ref{sample} inset~(a)]. The AB oscillations for this gate
voltage are displayed in the inset~(b) of Fig.~\ref{sample}. The
period of 1.1~mT is consistent with the ring radius of $r =
1~\mu$m. Fig.~\ref{2dtotal} shows a more detailed gate voltage
dependent magneto-conductance. For the displayed gate voltage
range, the change of carrier density is negligible. Here, two
regions can clearly be distinguished. Near the symmetry point,
there is no noticeable influence of the {\sc SO} coupling on the
position of the conductance oscillations [Fig.~\ref{B0}~(a)]. The
symmetry point of $V_{Gate} = -2.568$~V has been determined from
Fig.~\ref{2dtotal}.

For increasing $\Delta_{R}$ the situation is drastically
modified. Over a large range in gate voltage which corresponds to
$\Delta_{R}$ from zero up to $\sim~200~\mu$eV, the AB
oscillations show phase shifts and bifurcations with increasing
gate voltage (Fig.~\ref{2dtotal}). A repetitive sequence of
maxima and minima can be observed. Qualitatively, the observed
conductance fluctuations represent the expected interference
pattern due to phase differences for different spin orientation
of the partial waves propagating around the ring
(cf.~Eqs.~\ref{upup} - \ref{updown}). The phase differences are
caused by the change in strength of the {\sc SO} coupling and
thus the orientation of the total magnetic field with respect to
the ring plane. The oscillations are symmetric in magnetic field
and with respect to the applied gate voltage around the symmetry
value of $V_{Gate}=-2.568$~V. This behavior is expected for the
{\sc SO} coupling controlled dynamic part of the AC phase. In
contrast, the geometric phase contribution to Eq. (1) varies only
slowly with increasing SO interaction, so that one does not
anticipate a discernable signature of this effect for the gate
voltage range studied here.

To verify that the experimental data represent a direct
observation of the Aharonov-Casher phase in multichannel rings,
we compare them to numerical calculations within the
Landauer-B\"{u}ttiker (LB) formalism \cite{Souma04}. The
effective mass Hamiltonian for a
two-dimensional ring with Rashba {\sc SO} interaction and in a
perpendicular magnetic field, $B_z$, is given by:
\begin{eqnarray}\label{H2D}
\hat{H}_{r}=\frac{\mathbf{\hat{\pi}}^2}{2m_{eff}}+
\frac{\alpha}{\hbar}(\mathbf{\hat{\sigma}}\times
\mathbf{\hat{\pi}})_z + \hat{H}_Z
+\hat{H}_{conf}(r)+ \hat{H}_{dis}
\end{eqnarray}
where $\mathbf{\hat{\pi}}=\mathbf{\hat{p}}-e{\hat{\bf A}}$ and
$\hat{H}_Z=\frac{1}{2}g\mu_B\sigma_zB_z$. The first term is the
kinetic energy contribution, the second term corresponds to the
spin-orbit Rashba interactions, the third is the Zeeman
interaction, and $\hat{H}_{conf}$ and $\hat{H}_{dis}$ are the
confinement and disorder components of the Hamiltonian,
respectively. For an ideal 1D ring, the conductance in the
magnetic and Rashba fields can be found analytically
\cite{Meijer02,Nitta99,Frustaglia04,Molnar04}. However, our
actual experimental structures are not 1D and multichannel
effects have to be taken into account. Here, we use the
concentric tight-binding (TB) approximation to model the
multichannel rings extending the calculations in
Ref.~\cite{Souma04} to include not only the Rashba interactions
but also the the effect of magnetic field.

\begin{figure}[t]
\epsfig{figure=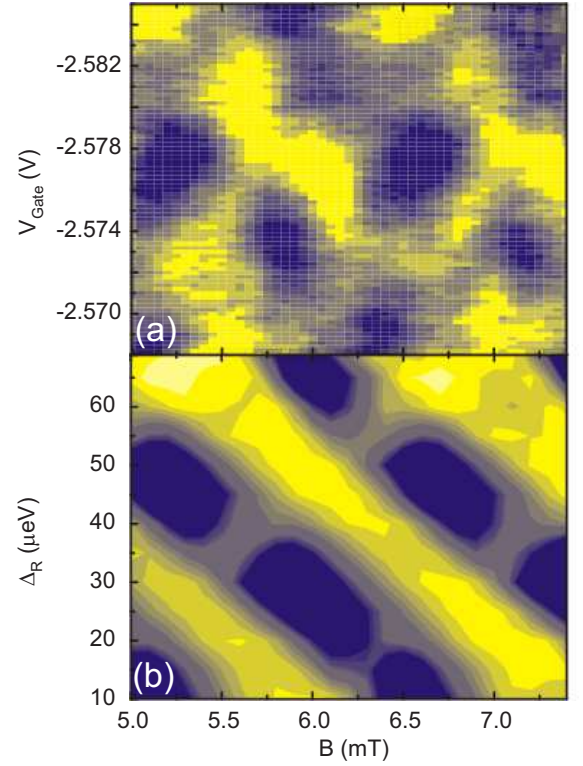,width=3.in}

\caption{When the SO interaction is modified via the gate
voltage, a shift of the conductance maxima (yellow) can be
observed due to the Aharonov-Casher phase (a). In figure (b), the
theoretical results for the conductance in a 6-channel ring as a
function of the Rashba energy and $B_{ext}$ are shown. The
scaling of the y-axis allows a direct comparison of the
experimental and theoretical data.}
\label{2Dfig}
\end{figure}

Within this approximation, the Hamiltonian becomes:
\begin{eqnarray}\label{TBH}
\hat{H}_{r}= && \sum_{n,m=1}^{N,M}\sum_{\sigma=\uparrow,\downarrow}
\epsilon_{nm}c^{\dag}_{nm,\sigma}c_{n,m,\sigma}\nonumber\\ &&
- \sum_{n,m=1}^{N,M}\sum_{\sigma=\uparrow,\downarrow} %
[t_{\Theta}^{n,n+1,m}e^{\frac{i\Phi r_m}{\Phi_0}}
c^{\dag}_{nm,\sigma}c_{n+1,m,\sigma'}+h.c.] \nonumber\\ &&
-\sum_{n=1}^N\sum_{m=1}^{M-1}\sum_{\sigma=\uparrow,\downarrow}[
t_r^{m,m+1,n} c^{\dag}_{nm,\sigma}c_{n,m+1,\sigma'}+h.c]\nonumber \\
\end{eqnarray}
where $n$ and $m$ designate the sites in the azimuthal ($\Theta$)
and radial directions (r), respectively;
$\varepsilon_{mn}=4t\sigma_0-\frac{1}{2}g\sigma_z\mu_{B}B_z$ is
the on-site energy where $t = \hbar^2/(2ma^2)$ and $a$ is the
lattice constant along the radial direction.
$t_{\Theta}^{n,n+1,m}$ and $t_r^{m,m+1,n}$ are the total nearest
neighbor hopping parameters in azimuthal and radial directions,
respectively defined in Ref.~\cite{Souma04}. In the Landau gauge,
the hopping parameter in azimuthal direction is modified through
the term $e^{\frac{i\Phi r_m}{\Phi_0}}$, where $\Phi=\pi B_za$,
$\Phi_0 =h/e$, and $r_m=r_1+(m-1)a$ is the radius of a ring with
$m$ modes in radial direction. The innermost ring radius
corresponds to $m=1$ while $m=M$ stands for the outermost ring
radius. We also assume that the lattice spacing along the
azimuthal direction in the outermost ring is the same as that in
the radial direction. The ring is attached to two semi-infinite,
paramagnetic leads that constitute reservoirs of electrons at
chemical potentials $\mu_1$ and $\mu_2$. The influence of
semi-infinite leads in the mesoscopic regime is taken into
account through the self energy term and the total charge
conductance is calculated as outlined in Ref.~\cite{Souma04}.

We use an effective electron mass of $m_{eff} =0.031 m_0$ and
effective g-factor $|g|=20$ in accordance with n-doped HgTe QW
parameters \cite{Zhang04}. From the experimental Hall
measurements, we set the electron concentration $n_{2D} = $
$1.5\times 10^{12}$ cm$^{-2}$. The Fermi energy is obtained from
the chosen carrier concentration assuming an infinite 2D gas. To
verify the experimentally determined zero value of {\sc SO}
interaction at $V_{Gate}(\Delta_{R}=0) = -2.568$~V we have
calculated the conductance of a 6-channel ring within the LB
formalism. Fig. \ref{B0}~(b) shows the calculated conductance as
a function of Rashba energy and magnetic field. We have found
that for small Rashba energies, less than $5~\mu$eV, the
interference pattern is almost unperturbed by {\sc SO}
interactions and displays the multichannel Aharonov-Bohm
oscillations, similar to the experimental data [see
Fig.~\ref{B0}~(a)]. Furthermore, a comparison of the experimental
and the theoretical data leads to the conclusion that a change in
the gate voltage of $10$~mV leads to a change of $35~\mu$eV in
$\Delta_R$, which is in good agreement with the results obtained
from the Shubnikov-de Haas oscillations.

To verify the existence of AC phase, we performed the conductance
calculations for much larger Rashba couplings. In the case of a
strictly 1D ring, the appearance of the AC phase leads to
periodic oscillations in conductance as a function of the Rashba
energy \cite{Frustaglia04,Molnar04}. In contrast, the
theoretically predicted interference pattern for conductance is
more complex for a multichannel ring. In this case, the
repetitive conductance minima and maxima move diagonally as a
function of {\sc SO} coupling, as can be seen in the theoretical
simulation [Fig.~\ref{2Dfig}~(b)]. For this calculation a ring
with six channels was assumed (in agreement with an experimental
estimate), where only one channel is conducting. This latter
assumption is corroborated by the experimental results. For more
than one conducting channels, the interference pattern would be
much more complex and smeared out. The distinct interference
pattern is a strong indication that there is only one conducting
channel, presumably because of impurities and imperfections in
the ring geometry. Additional non-conducting channels contribute
mainly to the non-oscillating background (cf. inset (b) of
Fig.~\ref{sample}), so that the number of non-conducting channels
plays only a minor role for the interference pattern. The
theoretical model reproduces the main features of experimental
data, i.e., the diagonal position of conductance maxima and
minima [Fig.~\ref{2Dfig}~(a)]. A more quantitative comparison of
the experimental and theoretical data is difficult and should
take into account incoherence effects as well as the change in
width of the ring which is cumbersome to estimate.

In conclusion, we have measured the transport properties of HgTe
ring structures with a continuously adjustable SO interaction. In
these structures, the AB-type magneto-conductance oscillations
exhibit significant phase changes when the Rashba SO splitting
energy is varied. A numerical analysis shows that these
fluctuations are a direct consequence of an Aharonov-Casher phase
contribution to the electronic wave function. Thus, the
experimental results provide the first direct observation of the
Aharonov-Casher effect.

\acknowledgements

We thank H.-A. Engel and D. Loss for useful discussions. The
financial support of the Deutsche Forschungsgemeinschaft (SFB~410)
and ONR (04PR03936-00) is gratefully acknowledged.

\bibliography{ring}

\end{document}